\documentclass[prl,aps,preprint,showpacs]{revtex4}

\usepackage{graphicx}

\begin{document}

\title{Understanding the Clean Interface Between Covalent Si and Ionic Al$_2$O$_3$}

  \author{H. J. Xiang}
  \affiliation{National Renewable Energy Laboratory, Golden, Colorado
    80401, USA}

  \author{Juarez L. F. Da Silva}
  \affiliation{National Renewable Energy Laboratory, Golden, Colorado
    80401, USA}

  \author{Howard M. Branz}
  \affiliation{National Renewable Energy Laboratory, Golden, Colorado 80401, USA}

  \author{Su-Huai Wei}
  \affiliation{National Renewable Energy Laboratory, Golden, Colorado
    80401, USA}

  \date{\today}

  \begin{abstract}
    The atomic and electronic structures of the (001)-Si/(001)-$\gamma$-Al$_2$O$_3$
    heterointerface are investigated by first principles total energy calculations
    combined with a newly developed ``modified basin hopping'' method.
    It is found that all interface Si atoms are 4-fold coordinated due
    to the formation of Si-O and unexpected
    covalent Si-Al bonds in the new abrupt interface model.
    And the interface 
    has perfect electronic properties in that the unpassivated
    interface has a large LDA band gap and no gap levels.
    These results show that it is possible to have clean
    semiconductor/oxide interfaces. 
  \end{abstract}

  \pacs{68.35.-p,73.20.-r,71.15.Nc,02.70.Uu}

  \maketitle
  Interfaces between semiconductors and metal oxides are playing
  increasingly important roles in advanced material science
  \cite{Forst2004,McKee1998,McKee2003}. In order to continue scaling
  electronic devices, a change from SiO$_2$ (with a dielectric
  constant $k$ about $3.9$) to high-$k$
  oxides has been
  proposed for the gate dielectric in future
  generation metal oxide semiconductor (MOS)
  technologies. 
  The key considerations for high-$k$ gate dielectrics include
  high dielectric
  constant, high band offsets (at least 1 eV) with respect to silicon,
  thermal stability, and minimization of electrical defects in the
  interface. In particular, the quality of the interface is important
  for both carrier mobility and device stability.
  However, control of the interface to the Si substrate remains a stubborn
  outstanding problem. 
  For example,  hafnium-based amorphous oxides has a bulk dielectric
  constant of $k\sim 22$ \cite{Ceresoli2006}, but its
  integration into the MOS gate stack poses substantial technological
  challenges \cite{Kingon2000}.
  Epitaxial growth of oxides could lead to more abrupt oxide-Si interfaces and consequently
  could offer solutions for the end of the roadmap. 
  Indeed, single crystal $\gamma$-Al$_2$O$_3$ ($k \sim 11$)
  thin films have been epitaxially grown by molecular beam epitaxy 
  on Si(001) substrates \cite{Merckling2006}. 
  Hence, Al$_2$O$_3$ could be a good candidate to be used directly as a gate
  oxide or as a thin buffer barrier when combined with high-$k$
  amorphous or epitaxial oxides.

  In another context, there have been some efforts in developing high
  quality crystalline silicon (c-Si) film on inexpensive foreign substrates
  such as oxides to reduce the Si material cost for terrestrial
  photovoltaic (PV)
  cells \cite{Teplin2006}. Previous attempts to grow single crystal Si
  on some oxides  such as CeO$_2$ failed due to the
  formation of SiO$_2$ \cite{Teplin2006}.
  Recently, Findikoglu {\it et al.} \cite{Findikoglu2005} demonstrated
  the growth of well-oriented Si thin films with high carrier mobility
  on $\gamma$-Al$_2$O$_3$ substrate.
  In addition, Al$_2$O$_3$ has been shown to passivate the c-Si surface
  efficiently for PV applications \cite{Hoex2006}.
  These results suggest $\gamma$-Al$_2$O$_3$ could be a good substrate for c-Si
  solar cell growth. Therefore, detailed knowledge of
  the Si/Al$_2$O$_3$ interface are vital.
  
  Although many experimental studies have examined the growth of
  (001) $\gamma$-Al$_2$O$_3$ on the Si (001) surface
  ($\gamma$-Al$_2$O$_3$/Si), the detailed interface structure remains
  unclear. Theoretically, Boulenc and Devos \cite{Boulenc2006}
  proposed an interface model for (001)-$\gamma$-Al$_2$O$_3$ grown on (001)-Si surface
  by incorporating a
  defective spinel model \cite{Menendez2005} of $\gamma$-Al$_2$O$_3$.
  To obtain an interface without gap states, they introduced
  passivating O atoms to replace Si-Al and Si-Si bonds
  with Si-O and Al-O bonds. 
  However, it is not clear if their proposed
  substrate and interface \cite{Boulenc2006} have the lowest total energies because only
  a few models were tested. 
  It is also not clear if a sharp
  gap-states free interface can exist because the large chemical and
  size difference between the semiconductor and the oxide.
  Therefore, it is desirable to obtain an improved microscopic understanding of the atomic
  and electronic structures of this important Si/Al$_2$O$_3$ interface. 

  In this Letter, we develop a new modified basin-hopping (BH) method to
  search for the
  lowest energy structure of the
  (001)-Si/(001)-$\gamma$-Al$_2$O$_3$ interface.
  It is found that the new interface structure presents not
  only Si-O bonds, but also  Si-Al bonds, with all Si atoms 4-fold
  coordinated. Our density functional calculation
  shows that the interface is semiconducting with a type-I band
  alignment. Our results support the use of $\gamma$-Al$_2$O$_3$ as a
  gate oxide or a substrate for the c-Si growth. 

  Our density functional theory calculations employed the frozen-core
  projector
  augmented wave method (PAW) \cite{PAW} encoded in the Vienna {\it ab  initio}
  simulation package \cite{VASP}, and the local density
  approximation (LDA). 
  We use a plane-wave cutoff energy of 400 eV, except for
  the search of interface structures by the BH method, where
  we use
  a soft O PAW pseudopotential with a cutoff energy of 212 eV.

\begin{figure}
   \includegraphics[width=6.5cm]{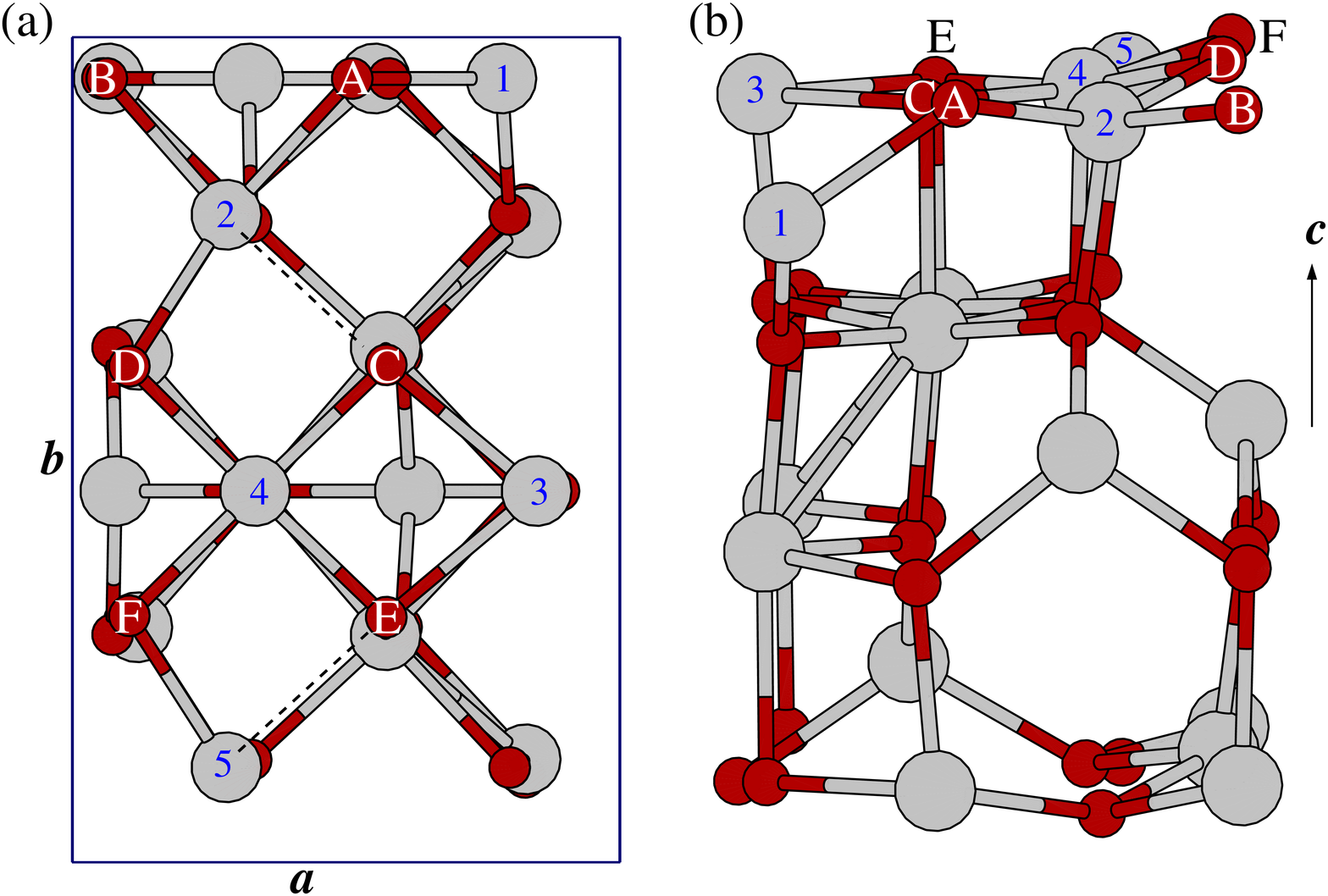}
   \caption{(Color online) (a) Top and (b) side view of the
     ($1\times1$) (001) $\gamma$-Al$_2$O$_3$ surface.
     Oxygen (small) atoms are indexed with capital letters and
     aluminum atoms (large) are indexed with numbers.}
   \label{fig1}
  \end{figure}

  As a prerequisite to build the Si/$\gamma$-Al$_2$O$_3$ interface, 
  understanding the $\gamma$-Al$_2$O$_3$ structure is necessary.
  Here, we adopt the bulk model 
  constructed by Krokidis {\it et al.}  \cite{Krokidis2001} (We
  hereafter refer to this bulk model as Krokidis model),
  which has a lower energy \cite{Digne2004} than traditional defective spinel models
  \cite{Menendez2005,Pinto2004}, and is consistent with
  experimental NMR and XRD results \cite{Zhou1991}. 
  Our test calculations also confirm the stability of the Krokidis
  model.  
  The Krokidis model has a centrosymmetric monoclinic structure (P2$_1$/m,
  No 11) with 1/4 four- and 3/4 six-fold
  coordinated Al atoms.
  Our LDA optimization results in the following parameters:
  $a=5.479$ \AA, $b=8.255$ \AA, $c=7.961$ \AA, and
  $\beta=90.645^\circ$.
  The lowest energy (001) $\gamma$-Al$_2$O$_3$ surface
  \cite{Digne2004} based on the Krokidis model has many inequivalent
  surface sites (Fig.~\ref{fig1}). 
  Here, oxygen atoms are indexed
  with capital letters and aluminum atoms with numbers
  \cite{Valero2006}.
  It is noted that there is a mirror symmetry plane which is perpendicular
  to $b$ and crosses Al 3 and 4. Therefore, the O at C (D) is
  equivalent to the O at 
  E (F), and the Al at 2 has the same environment as the Al at 5.
  All surface Al atoms  are pentacoordinated,
  except that  Al atom 1 is tetracoordinated and in a position
  slightly below the surface plane. All surface oxygens are
  tricoordinated if only Al neighbors within 2 \AA\ are counted.
  However, C and E oxygen atoms have an additional
  nearby Al atom besides the bonding Al atoms: i.e., the
  distance between Al 2 and oxygen C is 2.19 \AA. In this sense,
  C and E oxygen atoms are quasi-four-fold coordinated, as suggested in
  Fig.~\ref{fig1}(a) by the dashed lines. 
  Here, a four-layer (not counting the tetrahedral Al atoms)
  symmetric slab model is adopted. After relaxation, oxygen D (and F),
  oxygen A, and oxygen B
  move outward from the surface by about 0.3 \AA, 0.2 \AA, and 0.1 \AA,
  respectively. In contrast, oxygen C and E stay in the surface due to
  the strong binding with four neighboring Al atoms.
  It is noted that the surface is insulating due to the charge
  transfer from surface Al atoms to O dangling bonds. 
  
  We first examine the thermodynamic stability of the interface
  by calculating the enthalpy of two possible reactions \cite{Schlom2002}:
  \begin{equation}
    \begin{array}{ll}
      \frac{3}{2} \mathrm{Si} + \mathrm{Al}_2\mathrm{O}_3 \rightarrow   2
    \mathrm{Al} + \frac{3}{2}  \mathrm{Si}\mathrm{O}_2, \Delta H = 2.88 \quad
    \mathrm{eV} \\
    \mathrm{Si} + \frac{5}{3} \mathrm{Al}_2\mathrm{O}_3 \rightarrow  \frac{4}{3}
    \mathrm{Al} + \mathrm{Si}\mathrm{Al}_2{\mathrm O}_5, \Delta H =
    0.99 \quad
    \mathrm{eV}.
    \end{array}
  \end{equation}
  These positive reaction enthalpies indicate that the Si/Al$_2$O$_3$
  interface is thermodynamically stable, i.e., the formation of
  SiO$_2$ and silicate is unfavorable.

  The construction of the interface model is a nontrivial
  task. Usually, molecular dynamics simulations \cite{Broqvist2009} or intuition were
  employed for this purpose. It should be noted that molecular
  dynamics simulations
  gives different interface structures depending on 
  initial conditions, and
  it is almost impossible to guarantee that the
  constructed interface structure has the lowest interface energy. 
  And it is very hard to design a good interface structure between two
  totally dissimilar materials just from chemical intuition.
  Therefore, we develop a new modified
  BH method \cite{Wales1999} to determine the most stable interface structure.
  In conventional BH method, each BH run starts with a randomly chosen atomic
  configuration and is composed of a given number of Monte
  Carlo steps. In each of these, the starting configuration is first
  locally optimized to obtain an energy $E_1$. Then, each atom is subjected to a
  random displacement in each of its 
  Cartesian coordinates, and a new locally optimized structure is obtained with
  energy $E_2$. Here, $E_1$ and $E_2$ are the total energies
  from DFT calculations. If $\exp[(E_1-E_2)/k_BT]>r$,
  where $r$ is a random number between 0 and 1
  (Metropolis criterion), the new configuration is accepted
  (otherwise the old configuration is kept),
  and the process is iterated.
  The BH method has been widely used to
  search the global minimal structure of clusters \cite{Wales1999,Yoo2003,Pei2008,Barcaro2007}.
  However, to our best knowledge,  the BH method has not been employed to search
  for the interface structure between two surfaces.

\begin{figure}
   \includegraphics[width=6.5cm]{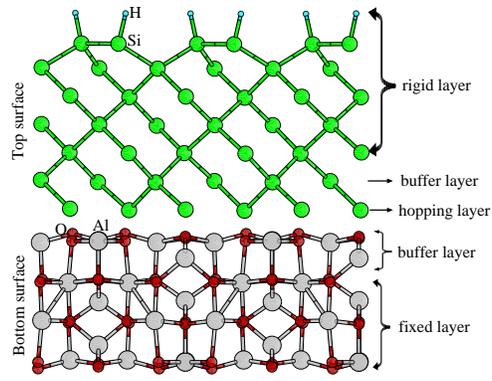}
   \caption{(Color online) The definition of various layers of the
     Si/Al$_2$O$_3$ interface in our modified BH method.}
   \label{fig2}
  \end{figure}

  In our newly developed modified BH method for finding lowest energy
  interface structures, we name the two slabs as ``top'' 
  and ``bottom'', respectively (see Fig.~\ref{fig2}). In the
  case of the Si/Al$_2$O$_3$ interface, the Si (Al$_2$O$_3$) (001) slab
  is the top (bottom) one. The Si slab has seven Si layers.
  The top Si layer forms Si dimers and is passivated by H atoms.
  For the top Si
  slab, we have a rigid layer, a buffer layer, and a hopping
  layer. The atoms in the rigid layer can translate as a rigid body
  but the internal structure is fixed \cite{fix}.
  The fixed layer is
  held in place and the buffer layer is
  allowed to relax during the optimization.
  In contrast, the atoms of the hopping layer move as
  in the usual BH method but are 
  restricted to the interface region.
  The typical value of the hopping distance of the BH
  simulation is about $1.5$ \AA.
  Our test calculations indicate that swapping a Si for an
  Al atom is energetically unfavorable by about 2 eV.
  Thus, the bottom Al$_2$O$_3$ slab is divided into two
  parts: a fixed layer and a buffer layer. 
  We note that the our modified BH method is rather general and 
  can be used to search for the lowest energy structure of other interfaces.

  Considering the lateral lattice contants
  of the (001) Si surface
  [$a(\mathrm{Si})=5.404$ \AA] and
  (001) $\gamma$-Al$_2$O$_3$ surface [$a(\mathrm{Al}_2\mathrm{O}_3)=
    5.479$ \AA,
  $b(\mathrm{Al}_2\mathrm{O}_3)  = 8.255$ \AA], the best lattice matching is
  achieved by connecting the ($1\times 3$) Si (001) surface with the 
  ($1\times 2$) (001) $\gamma$-Al$_2$O$_3$ surface.
  In this structure, the calculated lattice mismatch is  about 1.6\%.
  Here,
  the in-plane lattice constants of
  the supercell are fixed to be the theoretical lattice constants of
  bulk $\gamma$-Al$_2$O$_3$ because $\gamma$-Al$_2$O$_3$ has a large
  Young's modulus. 
  We perform several BH simulations for 200 steps with different initial
  coordinates (the relative position between the Si surface and the
  Al$_2$O$_3$ surface, and the atomic positions of the atoms of the
  hopping layer). 
  Finally, the lowest energy interface
  structure found from the BH simulations is refined by performing a
  full atomic relaxation of the whole system, including all atoms of
  the ``fixed'' Al$_2$O$_3$ layer and ``rigid'' Si layer.

\begin{figure}
    \includegraphics[width=6.5cm]{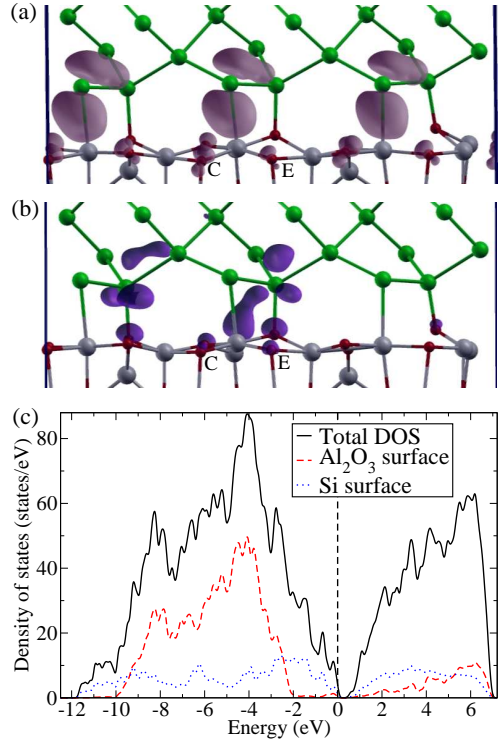}
    \caption{(Color online) Interface structure and isosurface plots of the partial charge
      density of (a) the topmost three HOMOs and (b) bottommost three
      LUMOs of the
      Si/Al$_2$O$_3$ interface.
      (c) DOS plot for the Si/Al$_2$O$_3$
      interface, 
      calculated with 0.1 eV broadening. The vertical dashed line
      denotes the top of the valence band. The partial DOSs of the Si
      and Al$_2$O$_3$ surfaces are also shown. }
    \label{fig3}
  \end{figure}

  The lowest energy interface structure that we find
  is shown in Fig.~\ref{fig3}. We can see that the
  dimer structure at the Si (001) surfaces is preserved as a result of
  the strong covalent Si-Si bond. We note that there is no dimer in
  the interface of the initial structure, while dimers are formed
  during the relaxation.
  At the interface, one Si atom of
  each dimer bonds with a three-fold coordinated O atom of the Al$_2$O$_3$ surface, whereas
  the other Si atom forms a bond with a four-fold coordinated Al atom. The Si-O and Si-Al
  bond lengths are about 1.8 \AA\ and 2.4 \AA, respectively.
  The O atoms bonded with Si move outward from the
  $\gamma$-Al$_2$O$_3$ surface in order to form bonds with 
  Si atoms. We find that oxygen C and E do not bond with Si atoms because
  it is unfavorable for them to move outward due to the strong
  binding with the fourth neighboring Al atom below the surface. 
  The binding energy between the Si surface and Al$_2$O$_3$
  surface is calculated to be 2.96 eV/supercell, which indicating the
  strong binding between the two surfaces. 
  It should be noted that
  there are some other nearly degenerate (within 20 meV/cell) interface structures with
  feature similar to that shown in Fig.~\ref{fig3}.
  In these metastable structures, other Al and O atoms are
  bonded with Si dimers.

  The DOS for the interface is shown in Fig.~\ref{fig3}(c). We can see
  that the system is semiconducting with an indirect band gap of 0.46
  eV. Remarkably, this value is larger than the LDA band gap (0.45 eV) of bulk
  Si. 
  The presence of the band gap is also consistent with the
  stability of the interface.
  The DOS plot shows a type-I
  band alignment between Si and Al$_2$O$_3$. 
  To compute an accurate band offset, we align the energy levels using the core
  levels \cite{Wei1998}. The calculated valence band offset is 2.40
  eV. The measured value between Si and
  $\alpha$-Al$_2$O$_3$ ranges from 2.90 eV to 3.75 eV
  \cite{Bersch2008}.
  The experimental valence band offset between Si and
  $\gamma$-Al$_2$O$_3$ is expected to have a similar value.
  The discrepancy between the experimental result and our
  theoretical value are due to the different LDA error for
  the covalent Si and ionic Al$_2$O$_3$ but the result is
  qualitatively correct \cite{Alkauskas2008}.  
  To gain insight into the electronic properties of the interface, 
  we show the partial  charge densities of the topmost three HOMOs
  and bottommost three LUMOs of the interface in Fig.~\ref{fig3}(a)
  and (b). It is clear that the HOMOs are mainly contributed by the
  directional covalent Si-Al bonds, and the LUMOs by the antibonding Si-O bonds.
  
\begin{figure}
    \includegraphics[width=6.5cm]{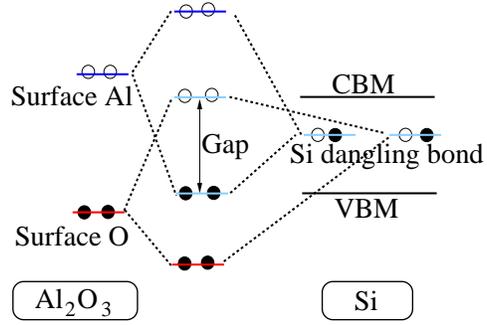}
    \caption{(Color online) Schematic illustration
      of the Si-Al and Si-O bond formation and gap opening in the
      Si/Al$_2$O$_3$ interface. The valence-band maximum (VBM) and
      conduction-band minimum (CBM) of bulk Si are also shown schematically.}
    \label{fig4}
  \end{figure}

  It is well known that each
  Si atom of the symmetric Si dimer of the Si (001) surface has one
  dangling bond. On the free Si (001) surface, the tilt of the Si dimer
  lifts the degeneracy of the Si dangling bonds and a band gap opens
  because of the charge transfer from the inward Si atom to outward Si
  atom.  
  In the case of the Si/Al$_2$O$_3$ interface, the band gap opening
  mechanism is totally different and much more efficient.
  As shown in Fig.~\ref{fig4}, the lone pair
  electrons of the surface O atom  interact with the dangling bond of
  the nearest neighbor Si atom, raising the level of
  the dangling bond. In contrast, the high-lying empty Al orbital
  hybridizes with the dangling bond of the neighboring Si atom,
  lowering the energy level of the Si orbital. 
  As a consequence, the Si atom bonded with the O atom transfers its
  dangling bond electron to the covalent Si-Al bond,
   and the interface has a large band gap.
  This bonding mechanism between Si and Al$_2$O$_3$ is
  consistent with the Bader charge analysis \cite{Bader}: 
  the Si atom bonded with Al
  gains about 0.25 electrons, whereas the Si atom bonded with O
  loses about 0.40 electrons. As a result, there is some small net charge transfer (0.15
  e/Si-dimer) from Si to Al$_2$O$_3$.

  To investigate the kinetic stability of the interface, we
  calculate the energy barrier of  the sliding of the Si surface on
  the Al$_2$O$_3$ surface. To find
  the transition state and energy barrier, we use the ``climbing
  image nudged elastic band'' method \cite{Henkelman2000}.
  We consider the sliding of the Si surface along the $b$ axis because
  the barrier of the sliding along other directions are expected to be
  larger due to the need to break all Si-O and Si-Al bonds.
  The final interface structure is obtained from the initial
  structure by sliding the Si surface along the $b$ axis by
  $b$(Al$_2$O$_3$)$/3$; the final state is almost degenerate with
  the initial state. In the transition state, there is some remaining
  bonding between the Si surface and Al$_2$O$_3$ surface:
  one Si-O bond and two Si-Al bonds. The energy barrier of the
  sliding is about 2.0 eV/supercell, which makes the Si/Al$_2$O$_3$ interface
  kinetically stable.

  In conclusion, we develop a general modified BH method to search for the 
  lowest energy structure of (001)-Si/$\gamma$-(001)-Al$_2$O$_3$
  interface.  
  It is found that the interface Si dimers have a favorable 4-fold
  coordination due to the formation of not only Si-O bonds, but also
  unexpected covalent Si-Al bonds. 
  Our study reveals that the Si/Al$_2$O$_3$ interface has the following
  attractive properties:
  (i) The interface is sharp and is semiconducting with a large LDA
  band gap;
  (ii) The band alignment between Si and $\gamma$-Al$_2$O$_3$ is type-I with
  both valence band offset and conduction band offset larger than 1.5 eV;
  (iii) The interface is thermodynamically and kineticly stable.
  Our results suggest that $\gamma$-Al$_2$O$_3$
  can be used as a gate dielectric in future
  MOS  technologies or a substrate for the growth of c-Si
  for solar cells.

  Work at NREL was supported by the U.S. Department of
  Energy, under Contract No. DE-AC36-08GO28308.

\clearpage

\clearpage

\clearpage

  \clearpage

\begin{thebibliography}{99}
\bibitem{fix} We modify the VASP code to relax a rigid body.
\end{thebibliography}
\end{document}